%Paper: 9203018
%From: romans@aristotle.jpl.nasa.gov (Larry Romans)
%Date: Sun, 8 Mar 92 13:44:51 PST

% % % % % % % % % % % % % % % % % % % % % % % % % % % % % % % % % % % % % % %

\def\al{\alpha} \def\be{\beta} \def\ga{\gamma} \def\de{\delta}
\def\ep{\epsilon}  \def\ka{\kappa} \def\la{\lambda}
 \def\th{\theta} \def\om{\omega} 
\def\Ga{\Gamma} \def\De{\Delta} \def\La{\Lambda} 
\def\Th{\Theta} \def\Om{\Omega} \def\Del{{\nabla\!{}}}
 
\def\pa{\partial} 
\def\fr#1#2{{{#1}\over{#2}}}
\def\1#1{\fr1{#1}} \def\2#1{\fr2{#1}} \def\3#1{\fr3{#1}}
\def\4#1{\fr4{#1}} \def\5#1{\fr5{#1}} \def\6#1{\fr6{#1}}
\def\7#1{\fr7{#1}} \def\8#1{\fr8{#1}} \def\9#1{\fr9{#1}}
\def\txs{\textstyle} 
\def\sst{\scriptstyle} 
\def\ft#1#2{{\txs{{#1}\over{#2}}}\.}
\def\fut#1#2{{\txs{{\raise1pt\hbox{$\sst #1$}}
  \over{\lower1pt\hbox{$\sst #2$}}}}}
\def\fraq#1#2{{\raise.3ex\hbox{$\sst #1$}\mkern-2mu/
  \mkern-2mu\lower.3ex\hbox{$\sst #2$}}}
\def\.{\mkern1mu} \def\x{\times}

\def\sb#1{\lower.3ex\hbox{${}_{\mkern-1mu#1}$}}

\def\ce#1{\centerline{{#1}}}
\def\ss{\smallskip}  \def\bs{\bigskip}
\def\na#1{\noalign{#1}} 
 \def\qq{\qquad}
\def\pren#1{\left({#1}\right)} \def\brak#1{\left[{#1}\right]}
\def\brac#1{\left\{{#1}\right\}} \def\bip#1{\big({#1}\big)}
\def\hb{\hfil\break} \def\ve{\vfill\eject}
\long\def\zilch#1zilch{}
\def\lyne#1#2{\baselineskip=\normalbaselineskip
  \multiply\baselineskip by #1 \divide\baselineskip by #2}
\def\boxit#1{\vbox{\hrule\hbox{\vrule\kern3pt
   \vbox{\kern3pt#1\kern3pt}\kern3pt\vrule}\hrule}}
\def\head#1{\goodbreak\vskip 0.5truein
  {\centerline{\bf #1}\par}\nobreak\vskip 0.25truein\nobreak}
\def\subhead#1{\goodbreak\vskip 0.15truein
  {\noindent{\bf #1}\hfil}\nobreak\vskip 0.1truein\nobreak}
\def\eef #1 {\llap{#1\enspace}\ignorespaces}
\def\reef{\head{References}\frenchspacing\parindent=0pt
  \leftskip=1truecm\parskip=4pt plus 2pt\everypar={\eef}}
\def\nofirstpageno{\nopagenumbers
  \footline={\ifnum\pageno>1\rm\hss\folio\hss\fi}}
\let\rawfootnote=\footnote
\def\footnote#1#2{{\rm\parskip=0pt\rawfootnote{#1}
  {\lyne11 #2\hfill\vrule height 0pt depth 6pt width 0pt}}}
\def\eg{{\it e.g.}} \def\ie{{\it i.e.}}
\def\npb #1 {Nucl. Phys. {\bf B{#1}} }
\def\plb #1 {Phys. Lett. {\bf B{#1}} }
\def\prd #1 {Phys. Rev. {\bf D{#1}} }
\def\prl #1 {Phys. Rev. Lett. {\bf{#1}} }
\def\cmp #1 {Commun. Math. Phys. {\bf{#1}} }
\def\ijmpa #1 {Int. J. Mod. Phys. {\bf A{#1}} }
\def\mpla #1 {Mod. Phys. Lett. {\bf A{#1}} }
\def\dj{\hbox{\kern.25em\raise.72ex\hbox{-}\kern-.58em d}}
\def\Dj{\hbox{\kern.05em\raise.31ex\hbox{-}\kern-.39em D}}
\def\z{\hbox{\kern.07em-\kern-.38em z}}
\def\Z{\hbox{\kern.2em\raise.3ex\hbox{-}\kern-.48em Z}}
\def\lqq{{,\kern-.1em,}} \def\rqq{{\kern-.1em``\kern.1em}}
\def\gsim{\mathrel{\raise.3ex\hbox{$>$}\mkern-14mu\lower.8ex\hbox{$\sim$}}}
\def\lsim{\mathrel{\raise.3ex\hbox{$<$}\mkern-14mu\lower.8ex\hbox{$\sim$}}}
\def\barr#1#2{{{#1}\mkern-1mu\llap{$\overline{\phantom{#2}}$}\mkern1mu}}
 \def\psib{{\barr\psi I}{}}
\def\epb{{\barr\ep\iota}} 
  
\def\SU{{\rm SU}} \def\Sp{{\rm Sp}} \def\SO{{\rm SO}} \def\U{{\rm U}}
\def\OSp{{\rm OSp}}  \def\cO{{\cal O}}
\def\wri#1{\immediate\write16{#1}}
\def\ciao{\ve\wri{}\wri{Best regards from L.R.}\wri{}\end}
\parindent=20pt \overfullrule=0pt \nofirstpageno

\def\RN{Reissner-Nordstr\o m} \def\BR{Bertotti-Robinson}
\def\hF{{\skew2\hat F}} \def\cD{{\cal D}} \def\hD{{\widehat\nabla\!{}}}
\def\gaa#1{\ga\sb{#1}}
\def\={\,=\,} \def\eqq{\,\equiv\,} \def\Fslash{{F\mkern-12mu/\mkern4mu}}

% % % % % % % % % % % % % % % % % % % % % % % % % % % % % % % % % % % % % % %

\lyne98
\rightline{{\tt hepth@xxx/9203018}}
\vskip 2truecm

\ce{\bf Supersymmetric, cold and lukewarm black holes}
\vskip .2truecm
\ce{\bf in cosmological Einstein-Maxwell theory}

\vskip 1.5truecm
\ce{L. J. Romans}
\vskip .5truecm
\ce{\sl Jet Propulsion Laboratory 301-150}
\ce{\sl California Institute of Technology}
\ce{\sl Pasadena, California 91109}
\vskip .2truecm
\ce{{\tt romans$\.\.$@$\.\.$aristotle.jpl.nasa.gov}}

\vskip 1.5truecm
\ce{ABSTRACT}
\bs
In flat space, the extreme Reissner-Nordstr\o m (RN) black hole
is distinguished by its coldness (vanishing Hawking temperature)
and its supersymmetry.
We examine RN solutions to Einstein-Maxwell theory with a
cosmological constant $\La$, classifying the cold black holes and,
for positive $\La$, the ``lukewarm" black holes at the same
temperature as the de Sitter thermal background.
For negative $\La$, we classify the supersymmetric solutions
within the context of $N=2$ gauged supergravity.
One finds supersymmetric analogues of flat-space extreme RN black holes,
which for nonzero $\La$ differ from the cold black holes.
In addition, there is an exotic class of supersymmetric solutions which
cannot be continued to flat space, since the magnetic charge becomes
infinite in that limit.

\vskip 1.0truecm
\lyne32

\head{1. Introduction}

Among the black-hole solutions to the Einstein or
coupled Einstein-Maxwell equations, the extreme
\RN\ (RN) black hole occupies a special position because
of its complete stability with respect to both classical
and quantum processes, including Hawking radiation,
permitting its interpretation as a soliton [1,2].
The extreme RN solution is also special in admitting
supersymmetry within the context of $N=2$ (ungauged) supergravity [2,3].
The connection between supersymmetry and stability has been
studied in detail [4--7], following Witten's proof of the positive
energy theorem for flat space making use of spinorial techniques [8].
Supersymmetry is also an integral element in demonstrating quantum
nonrenormalization theorems for RN black holes [9].

Analogues of the RN solutions to Einstein-Maxwell theory
with a cosmological constant $\La$ have been known for some time
[10,11].  This paper is essentially concerned with
identifying cosmological analogues of the extreme RN black holes,
with respect to two types of criterion:
stability against Hawking radiation, and supersymmetry.
Perhaps surprisingly, for nonzero $\La$ these criteria lead to
distinct varieties of solution.
Along the way, we encounter several special types of solution
with formally interesting properties: the ``ultracold" black holes
(in subsection 3.1) and the supersymmetric ``cosmic monopoles"
(subsection 4.3).

The organization of this paper is as follows.
The cosmological \RN\ solutions are reviewed in section 2 below,
along with the cosmological \BR\ solutions.
In section 3, applying general techniques for investigating the
thermodynamic properties of horizons [12--15],
we study the class of ``cold" black holes with vanishing Hawking
temperature.  For positive $\La$, in subsection 3.2 we characterize
the ``lukewarm" black holes whose temperature matches the de Sitter
background.

In section 4 we classify the supersymmetric RN solutions, explicitly
constructing the Killing spinors in all cases.
The natural framework for the analysis of supersymmetry
is the gauged version of $N=2$ supergravity [16], for which
$\La$ is necessarily negative.  One type of supersymmetric solution,
discussed in subsection 4.1, is a straightforward analogue of
the extreme RN black hole in flat space, whose supersymmetry [2--3]
is reviewed in subsection 4.2.  For nonzero $\La$, the minimal coupling
of the Maxwell field to the gravitini breaks the duality symmetry
between electric and magnetic fields, and in fact supersymmetry
singles out the purely electric solution.  The final class of
supersymmetric RN solutions, the ``cosmic monopoles" discussed
in subsection 4.3, have no flat-spcae analogue, since the magnetic
charge blows up in the formal limit to flat space.
It is interesting to note that the essentially local condition
of supersymmetry singles out a value for the magnetic charge
which automatically satisfies the Dirac quantization condition
(equation (4.8) below), obtained by global considerations.

Section 5 contains a discussion of our results and some of their
possible implications.
Finally, an appendix is devoted to the solution of a certain type
of constrained spinorial differential equation which arises at several
points in the analysis of supersymmetry.

\head{2. Solutions of cosmological Einstein-Maxwell theory}

The Lagrangian for Einstein-Maxwell theory with cosmological
constant $\La$ is given
by\footnote*{Our conventions are as follows:
Indices $m$, $n$, \dots\ are ``curved" world indices,
while $a$, $b$, \dots\ are ``flat" local Lorentz indices.
Specific curved indices are denoted by the coordinate
names $(t,r,\th,\phi)$, while numerical values
$(0,1,2,3)$ are reserved for flat indices.
We work in signature $\eta_{ab}$ $\equiv$ diag$\.(-,+,+,+)$,
with real gamma matrices satisfying $\{\gaa a,\gaa b\}$
= $2\.\eta_{ab}$.  Antisymmetrization is with unit weight,
\eg\ $\gaa{ab}$ $\equiv$ $\ga_{[a}\ga_{b\.]}$ $\equiv$
$\ft12[\gaa a,\gaa b]$.  The parity matrix $\gaa5$ $\equiv$
$\gaa{0123}$ is real and antisymmetric.}
$$
e^{-1}L\=-\ft14R\,+\,\ft14F_{mn}F^{mn}\,+\,\ft12\La\,.
\eqno(2.1)
$$
The field equations following from (2.1) read
$$
R_{mn}\=2\.F_{mp}F_n{}^{\,p}-\ft12g_{mn}F_{pq}F^{\,pq}
+\La g_{mn}
\eqno(2.2)
$$
and
$$
\pa_m\bip{eF^{mn}}\=0\,.
\eqno(2.3)
$$
Cosmological \RN\ solutions [10,11] can be described by metrics of the
stationary, spherically symmetric form
$$
ds^2\=-Vdt^2+\fr{dr^2}V+r^2\bip{d\th^2+\sin^2\th\,d\phi^2}\,,
\eqno(2.4)
$$
where the metric function $V(r)$ is taken to be a function of the
radial coordinate $r$ alone.
It will be convenient to define $U$ $\equiv$ $\sqrt{V}$ and
introduce the tetrad
$$
e_m{}^a\={\rm diag}\bip{\,U,U^{-1},r,r\sin\th\,}\,.
\eqno(2.5)
$$
The vector potential is taken to have nonvanishing components
$$
A_t\=\fr Qr\,,\qq A_\phi\=-H\cos\th
\eqno(2.6)
$$
with corresponding field strength
$$
F\sb{01}\=\fr Q{r^2}\,,\qq F\sb{23}\=\fr H{r^2}\,,
\eqno(2.7)
$$
describing electric charge $Q$ and magnetic charge $H$.
We assume that $Q$ and $H$ are not both zero, and define
$$
Z^2\eqq Q^2+H^2\,.
\eqno(2.8)
$$

It is straightforward to verify that the ansatz described above
provides a solution to the Einstein-Maxwell field equations
(2.2--3) when the metric function takes the form
$$
V(r)\=1-\fr{2M}r+\fr{Z^2}{r^2}-\ft13\La r^2\,.
\eqno(2.9)
$$
One sees from the behavior of the curvature invariants
$$\eqalignno{
R^2&\=16\La^2
&(2.10a)\cr
R_{mn}R^{mn}&\=4\pren{\fr{Z^4}{r^8}+\La^2}
&(2.10b)\cr
C_{mnpq}C^{mnpq}&\=\fr{48}{r^4}\pren{\fr Mr-\fr{Z^2}{r^2}}^2
&(2.10c)\cr
}$$
that such solutions possess a single
physical singularity located at the origin.
For large $r$, the metric is asymptotic to (anti-) de Sitter
space with cosmological constant $\La$.

We shall have occasion to deal with cosmological analogues of the
\BR\ (BR) solution to Maxwell-Einstein theory
[17,11].  Such a spacetime is locally the direct product of a
pair of maximally symmetric two-dimensional spacetimes,
with signatures $\eta\sb{\al\be}$ = diag$\.(-,+)$
and $\eta\sb{\al'\be'}$ = diag$\.(+,+)$, taking Greek and
primed Greek indices for the respective spaces.
The field strength is covariantly constant, with components
$$
F\sb{\al\be}\=E\ep\sb{\al\be}\,;\qq
F\sb{\al'\be'}\=B\ep\sb{\al'\be'}\,.
\eqno(2.11a,b)
$$
The cosmological Einstein equations become
$$
R\sb{\mu\nu}\=\La\sb1g\sb{\mu\nu}\,;\qq
R\sb{\mu'\nu'}\=\La\sb2g\sb{\mu'\nu'}
\eqno(2.12)
$$
with
$$
\La\sb1\=\La-(E^2+B^2)\,;\qq
\La\sb2\=\La+(E^2+B^2)\,.
\eqno(2.13)
$$
The values of $\La\sb1$ and $\La\sb2$ uniquely determine the
metrics upon the two-spaces.
For $\La\sb1$ positive (negative), $g\sb{\mu\nu}$ describes
two-dimensional (anti-) de Sitter space.
Meanwhile, for $\La\sb2$ positive (negative), $g\sb{\mu'\nu'}$
describes a two-sphere (hyperboloid).

\head{3. Cold and lukewarm \RN\ solutions}
\def\Zc{(Z^2)\sb{\rm crit}}

$>$From the forms of the curvature invariants (2.10$a$--$c$),
one sees that the only physical curvature singularity for
a metric of RN form is located at the origin; however, there will
in general be event horizons at the radii for which the metric
function $V(r)$ vanishes.
In asymptotically flat space ($\La=0$), for $Z^2<M^2$
there are two horizons, at radii $\rho_\pm$ = $M\pm\sqrt{M^2-Z^2}$
(see \eg\ [18]).  For nonzero $\La$, the condition $V(r)=0$, or
$$
-\ft13\La r^4+r^2-2Mr+Z^2\=0\,,
\eqno(3.1)
$$
is a quartic algebraic equation for $r$.  The standard closed-form
expressions for the four roots in terms of surds, being rather lengthy
and not especially illuminating, are omitted here.
As one would expect, if $\La$ is sufficiently
small compared to the local scales $M^{-2}$
(as in many cases of potential interest), the configuration of horizons
will display the same qualitative behavior as in flat space:
for $Z^2$ less that some critical value $\Zc$, there will be two
horizons at the roots $r=\rho_\pm$; these two horizons coalesce when
$Z^2=\Zc$, while for $Z^2>\Zc$ the double root bifurcates to a conjugate
pair of complex roots and one is left with a naked singularity.
The actual value of $\Zc$ can be determined by solving the cubic
equation in $Z^2$
$$
M^2-Z^2-\ft13\La\bip{27M^4-36M^2Z^2+8Z^4}-\ft{16}9\La^2Z^6\=0\,;
\eqno(3.2)
$$
we again omit the standard closed-form solution.
Defining a dimensionless parameter $\la$ $\equiv$ $M^2\La$,
for small $|\la|$ one finds
$$
\1{M^2}\,\Zc\=1\,+\,\ft13\la\,+\,\ft49\la^2\,+\,
\ft89\la^3\,+\,\cO(\la^4)\,.
\eqno(3.3)
$$
For $Z^2<\Zc$, one can parametrize the charge with the dimensionless
quantity $\De$ $\equiv$ $M^{-1}\sqrt{\Zc-Z^2}$.  The positions of
the two horizons are given for small $|\la|$ by
$$\eqalign{
\1M\,\rho_\pm&\=(1\pm\De)\,
+\,\ft16(2\pm\De)\bip{2\pm2\De+\De^2}\,\la\cr
&\qq+\,\ft1{72}\bip{96\pm204\De+224\De^2\pm140\De^3
+48\De^4\pm7\De^5}\,\la^2\,+\,\cO(\la^3)\,.
}\eqno(3.4)
$$

For positive cosmological constant, one finds an additional
root at a radius $b$ $\sim$ $\sqrt{3\La^{-1}}$ much farther
out than the local scales (again assuming small $|\la|$).
As discussed in [13], this root corresponds to an event
horizon constituting the ``outer edge" of the
de Sitter universe in the given coordinate system, and can be
treated on the same footing as the other roots.

Before examining specific configurations of horizons in more detail,
we turn to the issue of temperature for such horizons.
In general, one can deduce the Hawking temperature associated
with a horizon by considering the resolution of the coordinate
singularity in the Euclidean regime [15].
In the present situation, we examine the behavior of the metric (2.4)
near a root $\rho$ of $V(r)$ (either $r\gsim\rho$ or $r\lsim\rho$).
The surface gravity at the horizon [12] is given by
$$
\ka\=\ft12|V'(\rho)|\,.
\eqno(3.5)
$$
When $\ka$ is nonvanishing, in terms of a new
radial coordinate $x$ $\sim$ $\sqrt{2\ka^{-1}|r-\rho|}\.$,
the metric close to $x=0$ looks like
$$
ds^2\ \sim\ dx^2+x^2(i\ka\,dt)^2+\rho^2d\Om^2\,.
\eqno(3.6)
$$
The point $x=0$ is a perfectly regular point of the metric, if $it$
(\ie, $i\cdot t$) is regarded as an angular variable with period
$2\pi/\ka$.  The Hawking temperature for the horizon is given
by the inverse of this period [12--15]:
$$
T\=\fr\ka{2\pi}\,.
\eqno(3.7)
$$
For the RN solutions we consider, this gives
$$
T\=\1{4\pi}\big|V'(\rho)\big|
\=\1{4\pi\rho}\left|\,\,1-\fr{Z^2}{\rho^2}-\La\rho^2\,\right|\,.
\eqno(3.8)
$$
Taking $Z^2=0$ in (3.8) reproduces the result given in [19] for
the temperature of an uncharged cosmological black hole.
For general $Z^2<\Zc$ and small $|\la|$, (3.4) and (3.8) give
$$
T\=\fr\De{2\pi M(1+\De)^2}\,
-\,\fr{\De(14+26\De+19\De^2+5\De^3)}{12\pi M(1+\De)^3}\,\la
\,+\,\cO(\la^2)\,.
\eqno(3.9)
$$
for the temperature at the outer radius $\rho_+$.
The leading term here is the temperature of a general
RN black hole in flat space, as derived in [14].

\subhead{3.1 Cold and ultracold \RN\ solutions}

It is evident from (3.8) that horizons with vanishing
Hawking temperature are located at simultaneous roots
$\rho$ of both $V$ and $V'$; that is, at double roots of $V$.
When such a double root exists, the function $V(r)$ must take the form
$$
V\sb{\!\rm cold}(r)\=\pren{1-\fr\rho r}^2
\bip{1-\ft13\La(r^2+2\rho r+3\rho^2)}\,,
\eqno(3.10)
$$
with the corresponding critical relationships between
mass, charge, horizon radius and cosmological constant:
$$\eqalignno{
M&\=\rho\.(1-\ft23\La\rho^2)
&(3.11a)\cr
Z^2&\=\rho^2(1-\La\rho^2)\,.
&(3.11b)\cr
}$$
Eliminating $\rho$ from $(3.11a,b)$ leads to the relationship
(3.2) between the mass and the critical charge,
while inverting ($3.11a$) corresponds to the expansion
(3.4) in the case $\De=0$.

For any given $\La\leq0$, all positive values of $\rho$, $M$ and
$Z^2$ are admitted; given the value of any one of them, the
other two are fixed uniquely by (3.11$a,b$).

For positive $\La$, there is a maximum allowed radius
$\rho=\rho\sb{\rm max}=\La^{-1/2}$ at which the charge vanishes; the
resulting metric, known as the Nariai metric [20], is characterized by
$$
M\=\ft13\rho\,;\qq Z^2\=0\,;\qq\La\=\rho^{-2}
\eqno(3.12)
$$
and
$$
V\sb{\!\rm Nariai}(r)\=-\fr{r^2}{3\rho^2}\pren{1-\fr\rho r}^2
\pren{1+\fr{2\rho}r}\,.
\eqno(3.13)
$$
For $0<\rho<\rho_{\rm max}=\La^{-1/2}$, there is an extra
positive root $b$ to $V(r)$, given by
$$
b\=\sqrt{3\La^{-1}-2\rho^2}\ -\ \rho\,;
\eqno(3.14)
$$
note that for the range $0<\rho^2<\ft12\La^{-1}$, the extra horizon
at $b$ is outside the cold horizon at $\rho$, while for
$\ft12\La^{-1}<\rho^2<\La^{-1}$ it is inside.
Inverting (3.14), one can completely describe the solution in terms
of $\rho$ and $b$:
$$
M\=\fr{\rho(b+\rho)^2}{b^2+2\rho\.b+3\rho^2}\,;\qq
Z^2\=\fr{b\rho^2(b+2\rho)}{b^2+2\rho\.b+3\rho^2}\,;\qq
\La\=\3{b^2+2\rho\.b+3\rho^2}\,.
\eqno(3.15)
$$
The metric function (3.10) reads
$$
V\sb{\!\rm cold}(r)\=-\fr{r^2}{(b^2+2\rho\.b+3\rho^2)}
\pren{1-\fr\rho r}^2\pren{1-\fr br}\pren{1+\fr{2\rho+b}r}
\eqno(3.16)
$$
and the temperature at $b$ is
$$
T\sb b\=\fr b{2\pi(b^2+2\rho\.b+3\rho^2)}\pren{1-\fr\rho b}^2
\pren{1+\fr\rho b}\,.
\eqno(3.17)
$$
As mentioned earlier in this section, for small cosmological constant
($\La\ll\rho^{-2}$, or $b\gg\rho$), it is most natural to interpret the
faraway horizon at $b$ as the outer edge of the de Sitter universe [13].
Formally taking $\rho$ to zero in (3.15--17), the black hole disappears
and one reproduces the result
$$
T\sb{\rm dS}\=\1{2\pi}\sqrt{\fr\La3}
\eqno(3.18)
$$
for the background of thermal radiation in a pure de Sitter
cosmology [13].

A very special configuration is obtained when $b$ and $\rho$
coincide, leading to a triple root of $V(r)$.
The result is an ``ultracold" horizon at radius $\rho$ with zero
Hawking temperature, whose local structure is qualitatively different
from that encountered for the black holes discussed so far
(see (3.22) below).
The mass, charge and cosmological constant are related according to
$$
M\=\ft23\rho\,;\qq Z^2\=\ft12\rho^2\,;\qq\La\=\ft12\rho^{-2}\,,
\eqno(3.19)
$$
and the metric function reads
$$
V\sb{\!\rm ultracold}(r)\=-\fr{r^2}{6\.\rho^2}
\pren{1-\fr\rho r}^3\pren{1+\fr{3\rho}r}\,.
\eqno(3.20)
$$
This configuration simultaneously maximizes the values of
$M$ and $Z^2$ for any given positive value of $\La$.

For a generic zero-temperature black hole (for and $\La$),
the local structure of the metric at the horizon is no longer
described by (3.6), but by a cosmological \BR\ (BR) metric as discussed
in section 2 with
$$
\La\sb1\=2\La-\1{\rho^2}\,;\qq\La\sb2\=\1{\rho^2}\,.
\eqno(3.21a,b)
$$
For $\La=0$, one recovers the standard BR metric for the throat
of a flat-space extreme RN black hole (see \eg\ [18]).
Depending upon $\La$ and $\rho$, the spactime factor can be either
two-dimensional de Sitter or anti-de Sitter space;
the spacelike factor is clearly always a two-sphere of radius $\rho$

At the intermediate point for which $\La\sb1$ in ($3.21a$) vanishes,
one is dealing with the ultracold solution.
The throat in this case provides a sort
of formal interpolation between the cosmological BR spaces
with de Sitter and anti-de Sitter spacetime components.
The local structure at the throat is no longer approximated by
a BR metric, but rather by a metric of the form
$$
ds^2\ \sim\ dy^2\,-\,4(\rho/y)^6dt^2\,+\,\rho^2d\Om^2\,,
\eqno(3.22)
$$
appropriate for large $y$, where $y\sim\rho^{3/2}|r-\rho|^{-1/2}$.

\subhead{3.2 Lukewarm \RN\ solutions in de Sitter space}

Given that de Sitter space is awash with an isotropic background
of thermal radiation [13], it seems rather natural that a
black hole in a de Sitter background would be most comfortable
in a final configuration for which its (outer) horizon is at the
same temperature as the surrounding bath.
It was noted in [13] (see equation (5.3) there) that such a
configuration would avoid unpleasant branch cuts between coordinate
patches in the formulation of thermodynamics in black hole-de Sitter
spaces.  (However, no examples were available in [13], since charged
holes were not considered there.)

In this subsection, we find the ``lukewarm" RN solutions which
realize this state of affairs, that is, describing an outer
black hole horizon at radius $a$ and a de Sitter edge at radius $b$,
with the same Hawking temperature at $a$ and $b$.
In terms of the metric function $V(r)$, the algebraic problem is
$$
V(a)\=V(b)\=0\,;\qq V'(a)\=\pm V'(b)\,,
\eqno(3.23a,b)
$$
where the minus sign is appropriate, since there should be
no roots of $V$ between $a$ and $b$.
These conditions fix $V(r)$ to take the form
$$\eqalign{
V\sb{\!\rm lukewarm}(r)&\=-\fr{r^2}{(a+b)^2}\pren{1-\fr ar}
\pren{1-\fr br}\pren{1+\fr{(a+b)}r-\fr{ab}{r^2}}\cr\na\ss
&\=\pren{1-\fr{ab}{(a+b)r}}^2-\fr{r^2}{(a+b)^2}\,,\cr\na\ss
}\eqno(3.24)
$$
from which one reads off the mass, charge and cosmological constant:
$$
M\=\fr{ab}{(a+b)}\,;\qq Z^2\=\fr{a^2b^2}{(a+b)^2}\,;
\qq\La\=\3{(a+b)^2}\,.
\eqno(3.25)
$$
Remarkably, the charge and mass have the same simple relationship
$Z^2=M^2$ as for extreme RN black holes in flat space.
The mass $M$ has a ``reduced mass" form in terms of the two
radii $a$ and $b$.  The temperature at $a$ and $b$ is
$$
T\sb a\=T\sb b\=\fr{|\.b-a|}{2\pi(a+b)^2}\,.
\eqno(3.26)
$$
The black hole possesses an inner horizon at
$$
\rho_-\=\ft12\sqrt{a^2+6ab+b^2\,}\ -\ \ft12(a+b)\,.
\eqno(3.27)
$$
These expressions reduce to the proper results for empty de Sitter
space (taking $a$ $\to$ $0$) and for an extreme RN hole in
asymptotically flat space (taking $b$ $\to$ $\infty$).
When $a$ and $b$ coincide, one obtains one of the cold holes discussed
in the previous subsection, having $M=\ft12a$, $Z^2=\ft14a^2$,
$\La=\ft34a^{-2}$ and an extra horizon at $\rho_-=(\sqrt2-1)\.a$.

Expressed in terms of the mass and cosmological constant, the
inner and outer horizons for the black hole are at
$$
\rho_\pm\=\fr{2M}{1+\sqrt{1\mp 4M\sqrt{\La/3}}}
\ \sim\ M\,\pm\,M^2\sqrt{\La/3}\,+\cO(\La)\,,
\eqno(3.28)
$$
where of course $\rho_+$ is identified with $a$.
The de Sitter coordinate edge is at
$$
b\=\12\sqrt{\3\La}\pren{1+\sqrt{1-4M\sqrt{\La/3}}\,}\,,
\eqno(3.29)
$$
and the temperature (3.26) reads
$$
T\sb{\rho_+}\=T\sb b
\=\1{2\pi}\sqrt{\fr\La3\pren{1-4M\sqrt{\La/3}\,}}\,.
\eqno(3.30)
$$

\head{4. Supersymmetric \RN\ solutions}

We now turn to a classification of the supersymmetric RN solutions.
As mentioned in the introduction, it is appropriate to address
this issue within the context of $N=2$ gauged supergravity [16].
This theory describes a graviton,
a brace of real gravitini which we assemble into
a single complex gravitino field $\psi_m$ $\equiv$
$\psi_m^1+i\psi_m^2$, and a Maxwell vector field $A_m$
minimally coupled to the gravitini with strength $g$.
The Lagrangian is given by
$$\eqalign{
e^{-1}L&\=-\ft14R\,+\,\ft12\psib_m\ga^{mnp}\cD_n\psi_p
\,+\,\ft14F_{mn}F^{mn}
\,+\,\ft i8\bip{F+\hF}^{mn}\,\psib\sb p
\gaa{[m}\ga^{pq}\gaa{n]}\psi\sb q\cr
&\qq\qq-\,\ft12g\.\psib_m\ga^{mn}\psi_n\,-\,\ft32g^2\,.
}\eqno(4.1)
$$
Supersymmetry fixes the cosmological constant to be $\La=-3g^2$.
The Lorentz- and gauge-covariant derivative $\cD_m$
acting on spinorial objects is defined by
$$
\cD_m\=\Del_m-igA_m\,.
\eqno(4.2)
$$
in terms of the Lorentz-covariant derivative $\Del_m$
= $\pa_m+\ft14\om_m{}^{ab}\ga_{ab}$.
The supercovariant field strength is
$$
\hF_{mn}\=F_{mn}-{\rm Im}\,\bip{\psib_m\psi_n}\,.
\eqno(4.3)
$$
The theory is given in a first-order formalism for the gravity sector,
so that the curvature scalar $R$ appearing in (4.1) is a function of both
$e_m{}^a$ and $\om_m{}^{ab}$.
The spin connection is fixed by its own
algebraic equation of motion following from (4.1); one finds
$$
\om_{mab}\=\Om_{mab}-\Om_{mba}-\Om_{abm}
\eqno(4.4)
$$
where
$$
\Om_{mn}{}^a\=\pa_{\.[m}e_{n]}{}^a
-\ft12{\rm Re}\,\bip{\psib_m\ga^a\psi_n}\,.
\eqno(4.5)
$$
The action corresponding to (4.1) is invariant under the
$N=2$ supersymmetry transformation
$$\eqalignno{
\de e_m{}^a&\={\rm Re}\,\bip{\epb\ga^a\psi_m}
&(4.6a)\cr
\de\psi_m&\=\hD_m\ep
&(4.6b)\cr
\de A_m&\={\rm Im}\,\bip{\epb\psi_m}\,,
&(4.6c)\cr
}$$
where $\ep$ is an infinitesimal Dirac spinor, and
the supercovariant derivative is given by
$$
\hD_m\eqq\cD_m+\ft12g\.\ga_m
+\ft i4\hF_{ab}\ga^{ab}\ga_m\,.
\eqno(4.7)
$$

For an ansatz in which the gravitini vanish in the background, the
field equations following from (4.1) coincide with the Einstein-Maxwell
equations (2.2--3) with cosmological constant $\La=-3g^2$.
Therefore, the cosmological RN solutions discussed in section 2
(for negative $\La$) provide background solutions to gauged
$N=2$ supergravity.
A new feature arises when the RN solutions are considered in
this larger context: for a globally consistent interacting theory
the magnetic charge $H$ is subject to a Dirac quantization
condition, because of the minimal coupling between the gravitini
and the vector potential.  With our normalizations, this condition
states that the quantity
$$
n\eqq2gH
\eqno(4.8)
$$
is required to be an integer.

The supersymmetry admitted by a \RN\ solution to
gauged $N=2$ supergravity is characterized by
the space of solutions to the Killing spinor equation
$$
\hD_m\ep\=0\,,
\eqno(4.9)
$$
where the supercovariant derivative is evaluated in the
given background.  For the solution at hand,
the components of the supercovariant derivative read,
recalling that $Z^2$ $\equiv$ $Q^2+H^2$,
$$\eqalignno{
\hD_t&\=\pa_t+\1{2r}\pren{\fr Mr-\fr{Z^2}{r^2}+g^2r^2}\gaa{01}
+\ft12gU\.\gaa0-ig\.\fr Qr+\ft i4U\.\Fslash\.\gaa0
&(4.10a)\cr
\hD_r&\=\pa_r+\ft12gU^{-1}\.\gaa1+\ft14U^{-1}\.\Fslash\.\gaa1
&(4.10b)\cr
\hD_\th&\=\pa_\th-\ft12U\.\gaa{12}+\ft12gr\.\gaa2+\ft i4r\.\Fslash\.\gaa2
&(4.10c)\cr
\hD_\phi&\=\pa_\phi-\ft12U\sin\th\,\gaa{13}-\ft12\cos\th\,\gaa{23}
+\ft12gr\sin\th\,\gaa3+igH\cos\th+\ft i4r\sin\th\.\Fslash\.\gaa3
&(4.10d)\cr
}$$
where
$$
\Fslash\eqq F_{ab}\ga^{ab}\=\2{r^2}\bip{Q\.\ga^{01}+H\.\ga^{23}}\,.
\eqno(4.11)
$$

The basic integrability condition for (4.9) reads
$$
\Om_{mn}\ep\=0\,,
\eqno(4.12)
$$
where the set of operators $\Om_{mn}$ are defined by
$$\eqalign{
\Om_{mn}&\eqq\big[\,\hD_m,\hD_n\,\big]\cr
&\=\ft14C_{mn}{}^{ab}\ga_{ab}
\,+\,\ft i2\ga^{ab}\ga_{\,[n}\bip{\Del_{m]}F_{ab}}
\,+\,\ft i8gF_{ab}\bip{3\ga^{ab}\ga_{mn}+\ga_{mn}\ga^{ab}}\,.\cr\na\ss
}\eqno(4.13)
$$
One finds that each $\Om_{mn}$ factorizes into the product
$\Om_{mn}\=X_{mn}\Th$, where
$$
\Th\eqq U+gr\.\gaa1+\brac{\1r-\fr M{Z^2}}
\bip{i\.\gaa0\.Q-i\.\gaa{123}\.H}
\eqno(4.14)
$$
and the $X_{mn}$ are in general nonsingular, so than the set of
conditions (4.12) is satisfied if and only if $\ep$ is a zero mode
of $\Th$ itself.
The condition for the integrability condition (4.12) to admit
a solution is, therefore, that the parameters $g$, $M$, $Q$
and $H$ are chosen so that det$\.\Th$ vanishes identically
as a function of $r$.  A straightforward calculation gives
$$
{\rm det}\.\Th\=
\brak{\,1-2gH-\fr{(M^2-2gHMr)}{Z^2}\,}
\x\brak{\,1+2gH-\fr{(M^2+2gHMr)}{Z^2}\,}\,,
\eqno(4.15)
$$
which vanishes in three cases:
$$\eqalignno{
H\=0\,,&\qq Q^2\=M^2\,;
&(4.16a)\cr
g\=0\,,&\qq Z^2\=M^2\,;
&(4.16b)\cr
M\=0\,,&\qq H\=\pm\1{2g}\,.
&(4.16c)\cr
}$$
We note the happy circumstance that all three cases are in
satisfaction of the Dirac quantization condition (4.8).
For nonzero $g$, case (4.16$c$) in fact corresponds to the
lowest nonzero value of $|H|$ consistent with this condition.

For the two cases (4.16$a,c$) with nonzero $g$, the metric function
$V(r)=U(r)^2$ is always a strictly positive function of $r$
(see (4.17) and (4.29) below).  Thus the singularity at the origin
is not clothed by a horizon in these cases, in flagrant violation of
Penrose's principle of cosmic censorship [21].  We shall return to
this issue in the discussion section; for now, we forge ahead with
the analysis of supersymmetry.

It is important to note that solving the (first) integrability
condition (4.12) does not in itself
guarantee the existence of a solution to the original
Killing spinor equation (4.9); one must either consider a sufficient
set of higher integrability conditions [22] or
return to the original first-order equation.
We follow the latter course: in the following three subsections
we return to the Killing spinor equation, finding the
general solution to (4.9) for each set of parameters
(cases ($4.16a$--$c$)) for which
the integrability condition (4.12) admits a solution.

\subhead{4.1 Electric AdS extreme \RN\ solutions}

For $H=0$ and $Q^2=M^2$ (case ($4.16a$)), the metric function is
$$
U(r)\=\sqrt{\pren{1-\fr Mr}^2+g^2r^2}\,.
\eqno(4.17)
$$
Taking $Q=M$ (the case $Q=-M$ is related by a field redefinition),
we find that $\Th$ is proportional to an $r$-dependent
projection operator $\Pi$,
$$
\Pi\eqq\1{2U}\,\Th
\=\12\,+\,\1{2U}\brac{gr\.\gaa1-i\.\gaa0\pren{1-\fr Mr}}\,,
\eqno(4.18)
$$
so the integrability condition (4.12) becomes
$$
\Pi\ep\=0\,.
\eqno(4.19)
$$
Returning to the first-order equation (4.9), we note that with a little
algebra, the action of the supercovariant derivative upon
$\ep$ can be written
$$\eqalignno{
\hD_t\ep&\=\bip{\pa_t-\ft i2g}\ep
&(4.20a)\cr
\hD_r\ep&\=\pren{\pa_r-\fr M{2r(r-M)}
+\fr{g(r-2M)}{2U(r-M)}\,\gaa1}\ep
&(4.20b)\cr
\hD_\th\ep&\=\bip{\pa_\th-\ft i2\gaa{012}}\ep
&(4.20c)\cr
\hD_\phi\ep&\=\bip{\pa_\phi-\ft12\cos\th\,\gaa{23}
+\ft i2\sin\th\,\gaa{013}}\ep\,,
&(4.20d)\cr
}$$
assuming that $\ep$ satisfies the necessary condition (4.19).

Since $\hD_r$ and $\Pi$ both commute with $\hD_t$, $\hD_\th$ and $\hD_\phi$,
one can address the temporal and angular equations
($4.20a,c,d$) of the Killing spinor equation
independently of the radial equations (4.19) and ($4.20b$).
One finds that
$$\eqalign{
\ep\.(t,r,\th,\phi)&\=\exp\bip{\ft i2gt}\exp\bip{\ft i2\gaa{012}\,\th}
\exp\bip{\ft12\gaa{23}\,\phi}\ep\.(r)\cr
&\=\exp\bip{\ft i2gt}\bip{\cos\ft12\th+i\.\gaa{012}\.\sin\ft12\th}
\bip{\cos\ft12\phi+\gaa{23}\.\sin\ft12\phi}\ep\.(r)
}\eqno(4.21)
$$
where the problem has been reduced to finding the spinors $\ep\.(r)$
satisfying ($4.20b$) as well as the projection condition (4.19).
The solution to this type of problem is provided in the appendix;
in the present situation, we find
$$
\ep\.(r)\=\pren{\sqrt{\.U(r)+gr\,}\,
+\,i\.\gaa0\,\sqrt{\.U(r)-gr\,}\,}P(-\gaa1)\,\ep\sb0\,,
\eqno(4.22)
$$
completing the description of the general solution to the Killing
spinor equation (4.9) in case ($4.16a$).
Note that because of the projection $P(-\gaa1)$,
the solution space is reduced from four to two (complex) dimensions.

In the pure AdS background, the algebra of supercharges is given
by the full $N=2$ AdS algebra $\OSp(2\.|\.4,R)$, with bosonic component
$\SO(2)\x\Sp(4,R)$ $\cong$ $\U(1)\x\SO(3,2)$, where $\U(1)$
corresponds to the Maxwell gauge invariance and $\SO(3,2)$
is the anti-de Sitter isometry algebra.
Introducing the supersymmetric RN solution described here
breaks the supersymmetry algebra to $\SU(2\.|\.1)$, whose bosonic
component is $\SU(2)\x\U(1)$, corresponding to independent
spatial rotations and time translations.
Note that the algebra $\U(1)$ of time translations
is compact; this is because AdS is the universal cover of a base
space identified under the transformation $t$ $\to$ $t+2\pi/g$
[23].  The Killing spinor (4.21) is antiperiodic under this shift
(just as $\ep$ $\to$ $-\ep$ under $\phi$ $\to$ $\phi+2\pi$).

\subhead{4.2 Flat-space extreme \RN\ solutions}

For the case ($4.16b$) with vanishing gauge coupling $g$, we are dealing with
the ungauged $N=2$ theory in flat space.  The supersymmetry
in this case has already been noted in the literature [2,3].
We include a brief discussion here, with an explicit construction
of the Killing spinors, in the interest of completeness
and ease of comparison with the other cases.

For $g=0$ and $Z^2=M^2$, we introduce the parametrization
$$
Q\=M\cos\al\,,\qq H\=M\sin\al\,.
\eqno(4.23)
$$
In this case the metric function takes the simple form
$$
U(r)\=1-\fr Mr\,.
\eqno(4.24)
$$
We first note that for the purely electric hole ($\al=0$),
one can simply take the results of the previous
subsection, formally setting the gauge coupling $g$ to zero.
One finds that the general Killing spinor is
$$
\ep\.(r,\th,\phi;0)\=\sqrt{\.U(r)}\,\exp\bip{\ft12\gaa{12}\.\th}
\exp\bip{\ft12\gaa{23}\.\phi}\,P(i\.\gaa0)\,\ep\sb0\,,
\eqno(4.25)
$$
where for convenience we have redefined the constant spinor
$\ep\sb0$ with respect to the one appearing in (4.22).
We have also noted that the generator $\ft i2\gaa{012}$ of $\th$
rotations in (4.21) can be replaced by the ``intuitive" generator
$\ft12\gaa{12}$, because of the presence of the projector
$P(i\.\gaa0)$.  The analogous replacement for $g\neq0$ would lead to a
rather complicated $r$-dependent generator, as a result of the
more complicated form of the projector $\Pi$ given in (4.18).

The simplest way to exhibit the Killing spinors
for nonzero values of $\al$ is by applying a duality transformation
to the result (4.25).  One easily verifies that, for $g=0$,
the supercovariant derivative for general $\al$ is given by
$$
\hD_m(\al)\=\exp\bip{-\ft12\gaa5\.\al}
\,\hD_m(0)\,\exp\bip{+\ft12\gaa5\.\al}\,.
\eqno(4.26)
$$
With this observation it is rather simple to construct the Killing
spinors for $\hD_m(\al)$ from those of $\hD_m(0)$.  One finds
$$
\ep\.(r,\th,\phi;\al)\=\sqrt{\.U(r)}\,\exp\bip{\ft12\gaa{12}\.\th}
\exp\bip{\ft12\gaa{23}\.\phi}\,P_\al\,\ep\sb0\,,
\eqno(4.27)
$$
where $P_\al$ is a projection operator commuting with the angular
factors, defined by
$$\eqalign{
P_\al&\eqq\exp\bip{-\ft12\gaa5\.\al}
\,P(i\.\gaa0)\,\exp\bip{+\ft12\gaa5\.\al}\cr
&\=\ft12\bip{1+i\.\gaa0(\cos\al+\gaa5\.\sin\al)}\,.
}\eqno(4.28)
$$
This provides an explicit construction of the Killing spinors
for a general (flat-space) extreme RN black hole.

\subhead{4.3 Exotic AdS solutions (``cosmic monopoles")}

We now consider the case ($4.16c$), with vanishing mass parameter
($M=0$) and magnetic charge $H=1/(2g)$;
the other sign for $H$ can be obtained by a field redefinition.
The metric function becomes
$$
U(r)\=\sqrt{\pren{gr+\1{2gr}}^2+\fr{Q^2}{r^2}}\,,
\eqno(4.29)
$$
where the electric charge $Q$ is arbitrary.
This case clearly has no flat-space limit, since
$H$ and $U(r)$ blow up as $g$ is taken to zero.
One might call such a solution a ``cosmic monopole"
(or cosmic dyon, for nonzero $Q$), since the characteristic
scales for size and magnetic charge are given by the overall
AdS cosmological distance scale ($\sim1/g$).
It is evidently inappropriate to study such solutions as local,
perturbative phenomena, that is, on scales small compared to $1/g$.
In contrast, for the RN solution of subsection 4.1,
the local scales are fixed by a parameter $M$ independent of,
and conceivably much smaller than, the global scale $1/g$
(which can formally be taken to infinity, as in subsection 4.2).

Considering the Killing spinor integrability condition (4.12),
we find that in this case the matrix $\Th$ of (4.14) is not itself
proportional to a projection operator, but turns out to be
twice as potent.  Consider the identities
$$\eqalignno{
\12\brak{\,U-gr\.\gaa1-\1r\pren{i\.\gaa0\.Q-\1{2g}\.i\.\gaa{123}}}
\Th&\=P(i\.\gaa{23})
&(4.30a)\cr
\1{4Ugr}\,\,\gaa1\,
\brak{\,U+gr\.\gaa1-\1r\pren{i\.\gaa0\.Q-\1{2g}\.i\.\gaa{123}}}
\Th&\=\Pi\,,
&(4.30b)\cr
}$$
where the notation $P(i\.\gaa{23})$ is defined in (A.9), and
$$
\Pi\eqq\12\,+\,\1{2U}\brac{\pren{gr+\1{2gr}}\gaa1+i\.\gaa0\,\fr Qr}
\eqno(4.31)
$$
is a second projection operator commuting with $P(i\.\gaa{23})$.
It is clear from ($4.30a,b$) that the integrability condition
$\Th\ep=0$ implies the pair of independent conditions
$$
P(i\.\gaa{23})\ep\=0\,;\qq\Pi\ep\=0\,.
\eqno(4.32a,b)
$$
In fact, $\Th\ep=0$ is completely equivalent to ($4.32a,b$),
since $\Th$ itself is a linear combination of these two projectors:
$$
\Th\=2\.U\,\Pi-\1{gr}\.\gaa1\,P(i\.\gaa{23})\,.
\eqno(4.33)
$$

The action of $\hD_m$ upon $\ep$ simplifies to a
remarkable degree modulo the two conditions ($4.32a,b$), yielding
$$\eqalignno{
\hD_t\ep&\=\pa_t\ep
&(4.34a)\cr
\hD_r\ep&\=\pren{\pa_r-\1{2r}-\fr gU\,\gaa1}\ep
&(4.34b)\cr
\hD_\th\ep&\=\pa_\th\ep
&(4.34c)\cr
\hD_\phi\ep&\=\pa_\phi\ep\,.
&(4.34d)\cr
}$$
The Killing spinors are therefore functions of $r$ alone.
Noting that the independent condition (4.32$a$) can be solved
trivially, the appendix provides the solution to the constrained
radial differential equation; one finds
$$
\ep\.(r)\=\pren{\sqrt{\.U(r)+gr+1/(2gr)}
\,-\,i\.\gaa0\,\sqrt{\.U(r)-gr-1/(2gr)}\,}
P(-\gaa1)P(-i\.\gaa{23})\,\ep\sb0\,.
\eqno(4.35)
$$
The arbitrary constant spinor $\ep\sb0$ is now subjected to a
double projection, reducing the original four components to one.
The lack of angular dependence for $\ep$ can be understood
in terms of group theory, since a one-dimensional representation of
the spin group SU(2) must of necessity transform as a singlet
(\ie, not transform at all) under rotations.

The purely magnetic case ($Q=0$) is in some formal respects
analogous to the flat-space extreme RN solution
examined in the preceding subsection.
The metric function is a rational function,
$$
U(r)\=gr+\1{2gr}\,,
\eqno(4.36)
$$
and the Killing spinor (4.35) again reduces to a simple product of
$\sqrt{\.U(r)}\,$ with a constant spinor:
$$
\ep\.(r)\=\sqrt{\.2\.U(r)}\,P(-\gaa1)P(-i\.\gaa{23})\,\ep\sb0\,.
\eqno(4.37)
$$
It is curious to note that in this case $U(r)$ and $\ep\.(r)$
are invariant under the spatial inversion
$$
r\to\1{2g^2r}\,.
\eqno(4.38)
$$
However, the full metric is not invariant; the spatial component
suffers a conformal rescaling.

\head{5. Discussion}

Within the class of electrically and magnetically charged
\RN\ solutions of Einstein-Maxwell theory with a cosmological
constant $\La$, we have classified the ``cold" black holes
with vanishing Hawking temperature (for general $\La$),
the ``lukewarm" black holes at the same temperature as the
thermal background (for $\La>0$), and the supersymmetric holes
(for $\La\leq0$).

In de Sitter space, the lukewarm black holes provide
natural candidates for equilibrium configurations analogous to the
extreme RN black holes in flat space.
Of course, our thermodynamic reasoning in this matter is rather
heuristic, and is no substitute for a detailed dynamical analysis.
It may be relevant that the lukewarm configurations
are precisely those which satisfy natural analyticity requirements
for the consistent global formulation of finite temperature field
theory in black hole-de Sitter spaces [13].

There is no intrinsic background temperature in anti-de Sitter space,
hence no concept of ``lukewarmth."
{}From experience with background solutions for supergravity
theories (see \eg\ [4--6]), one is led to expect the natural stable
ground states to be provided by the supersymmetric solutions.
It is interesting to note that the supersymmetric analogues of the
flat-space extreme RN solutions exactly preserve the relationship
$M^2=Z^2$, as do the lukewarm black holes.
Thus, the same simple condition could in principle characterize the
stable vacua of \RN\ type, for any value of the cosmological constant.

A significant aspect of the supersymmetric RN solutions
in AdS is the presence of naked singularities, violating Penrose's
principle of cosmic censorship [21].
The naked singularity for overextreme ($Z^2>M^2$) RN solutions appears
to be unstable in flat space, shedding charge until the extreme case
is reached and a horizon appears [24].
It would be interesting to see an analogous dynamical study in AdS.
We remark that if the hole continues to shed charge until a horizon
is formed, the result will be a cold black hole (with $Z^2$
slightly less than $M^2$ for small $|\La|$; see equation (3.3)).
On the other hand, if the process terminates when the supersymmetric
configuration is reached, one must ensure that the background makes
sense physically; in particular, the Cauchy problem must remain
well-defined.  One might hope that sufficiently careful attention to
boundary conditions at the singularity could resolve this issue.
This would be somewhat reminiscent of the need to impose rather delicate
boundary conditions at spatial infinity, in order to have a well-defined
Cauchy problem in anti-de Sitter space itself [23,5,25,26].

To directly address the stability of the supersymmetric solutions
discussed herein, one could attempt to modify
existing stability proofs for black holes [27], for supergravity
theories [4--7] and for anti-de Sitter backgrounds [5--7,28]
to treat the present case.  Again, the most critical issue involves
the proper boundary conditions at the singularity.
For the cosmic monopoles of subsection 4.3, we note
that the Dirac quantization condition (4.8) guarantees that among
the configurations with minimal magnetic charge $H=1/(2g)$, at least
one will be stable for purely topological reasons
(similarly for $H=-1/(2g)$).

In closing, we mention a few possible directions for extending this work.
It would be interesting to see a detailed formulation of
thermodynamics in the lukewarm black hole-de Sitter cosmologies;
this also provides a new laboratory for studying subtleties in
the description of scattering processes involving black holes [29].
One wonders whether considering charged configurations might affect
the early history of inflationary scenarios with de Sitter
phases (see \eg\ [30] for a review) or the perdurance properties
of de Sitter space [31].
Cosmological analogues could be considered for black-hole solutions in
field theories incorporating scalar fields (see \eg\ [32]).
Finally, it is perhaps not inconceivable that aspects of \RN\ solutions
discussed here might somehow be reflected in the structure of
certain conformal field theories (see \eg\ [33]) which admit an
interpretation in terms of charged black holes.

\head{Appendix. Solution to a certain spinorial differential equation}
\def\A{{\rm A}}

In addressing the Killing spinor equation for the backgrounds
studied in subsections 4.1 and 4.3, we encounter a certain type
of radial differential equation for a spinor $\ep(r)$ subject
to an $r$-dependent projection constraint.
In this appendix we present the solution to such problems in a
somewhat generalized context.

Suppose that $\Ga\sb1$ and $\Ga\sb2$ are constant operators satisfying
$$
\bip{\Ga\sb1}^2\=\bip{\Ga\sb2}^2\=1\,,\qq
\Ga\sb1\Ga\sb2\=-\Ga\sb2\Ga\sb1\,,
\eqno(\A.1)
$$
and that $x(r)$ and $y(r)$ are functions with $x^2+y^2=1$;
we assume that $y\neq0$.
The operator
$$
\Pi(r)\eqq\ft12\bip{1+x(r)\.\Ga\sb1+y(r)\.\Ga\sb2}
\eqno(\A.2)
$$
is then a projection operator for any $r$.
We consider the differential equation
$$
\eta'(r)\=\bip{a(r)+b(r)\.\Ga\sb1+c(r)\.\Ga\sb2}\eta(r)
\eqno(\A.3)
$$
subject to the constraint
$$
\Pi\eta\=0\,.
\eqno(\A.4)
$$
Since we assume $y\neq0$, (A.4) can be used to
absorb the $\Ga\sb2\eta$ term on the right-hand side of (A.3)
into the other two terms.  Therefore, without loss of generality
we take $c(r)=0$.

The condition for integrability of this system reads
$$
x'+2by^2\=0\,.
\eqno(\A.5)
$$
When this condition is satisfied (as one verifies in the cases
of interest), the general solution, given in terms of an arbitrary
constant spinor $\eta\sb0$, reads
$$
\eta(r)\=\bip{u(r)+v(r)\Ga\sb2}P(-\Ga\sb1)\,\eta\sb0
\eqno(\A.6)
$$
where
$$
u\=\sqrt{\fr{1+x}y}\,\exp w\,\qq
v\=-\sqrt{\fr{1-x}y}\,\exp w
\eqno(\A.7)
$$
and the function $w(r)$ is simply the antiderivative of $a(r)$:
$$
w(r)\=\int^r\!\!a(r')\,dr'\,.
\eqno(\A.8)
$$
The arbitrary additive constant for $w$ corresponds to
an overall scaling for $\eta$.

We have introduced the notation
$$
P(\Ga)\eqq\ft12\bip{1+\Ga}\,,
\eqno(\A.9)
$$
defined for any operator $\Ga$ with $\Ga^2\=1$,
to denote the projection operator onto the $+1$ eigenspace of $\Ga$.
Clearly, only the projection $P(-\Ga\sb1)\eta\sb0$ of the arbitrary
spinor $\eta\sb0$ actually contributes to the solution.

\head{Acknowledgments}

I am grateful to Gary Gibbons for several useful comments.
This work was supported by a National Research Council
Research Associateship.

\lyne98
\reef

[1] P. H\'aj\'\i\v cek, \npb 185 (1981) 254.

[2] G. W. Gibbons, in ``Proceedings of the Heisenberg Symposium,"
eds. P. Breitenlohner and H. P. Durr (Springer, Berlin, 1982).

[3] P. C. Aichelberg and R. G\"uven, \prd 24 (1981) 2066;
\prd 27 (1983) 456; \prl 51 (1983) 1613.

[4] G. W. Gibbons and C. M. Hull, \plb 109 (1982) 190.

[5] P. Breitenlohner and D. \Z. Freedman, \plb 115 (1982) 197;
Ann. Phys. {\bf144} (1982) 249.

[6] G. W. Gibbons, C. M. Hull and N. P. Warner, \npb 218 (1983) 173.

[7] C. M. Hull, \cmp 90 (1983) 545.

[8] E. Witten, \cmp 80 (1981) 381.

[9] R. Kallosh, ``Supersymmetric Black Holes," Stanford preprint
SU-ITP-92-1.

[10] F. Kottler, Annalen Physik {\bf56} (1918) 410.

[11] D. Kramer, H. Stepani, E. Herlt and M. MacCallum,
``Exact Solutions of Einstein's Field Equations" (Cambridge, 1980).

[12] S. W. Hawking, \cmp 43 (1975) 199.

[13] G. W. Gibbons and S. W. Hawking, \prd 15 (1977) 2738.

[14] G. W. Gibbons and M. J. Perry, Proc. R. Soc. Lond. {\bf A358}
(1978) 467.

[15] S. W. Hawking, \prd 18 (1978) 1747.

[16] A. Das and D. \Z. Freedman, \npb 120 (1977) 221;
\hb E. S. Fradkin and M. A. Vasiliev, Lebedev Institute preprint
N 197 (1976).

[17] B. Bertotti, Phys. Rev. {\bf116} (1959) 1331;
\hb I. Robinson, Bull. Acad. Polon. Sci. {\bf7} (1959) 351.

[18] C. W. Misner, K. S. Thorne and J. A. Wheeler, ``Gravitation"
(W. H. Freeman, New York, 1973).

[19] S. W. Hawking and D. N. Page, \cmp 87 (1983) 577.

[20] H. Nariai, Sci. Rep. T\^ohoku. Univ., I., {\bf35} (1951) 62.

[21] R. Penrose, Rev. Nuovo Cimento {\bf1} (1969) 252.

[22] P. van Nieuwenhuizen and N. P. Warner, \cmp 93 (1984) 277.

[23] S. W. Hawking and G. F. R. Ellis, ``The Large Scale Structure
of Space-time" (Cambridge, 1973).

[24] G. W. Gibbons, \cmp 44 (1975) 245.

[25] S. J. Avis, C. J. Isham and D. Storey, \prd 18 (1978) 3565.

[26] S. W. Hawking, \plb 126 (1983) 175.

[27] G. W. Gibbons, S. W. Hawking, G. T. Horowitz and M. J. Perry,
\cmp 88 (1983) 295.

[28] L. F. Abbot and S. Deser, \npb 195 (1982) 76.

[29] J. Preskill, P. Schwarz, A. Shapere, S. Trivedi and F. Wilczek,
``Limitations on the statistical description of black holes,"
Princeton preprint IASSNS-HEP-91/34.

[30] K. A. Olive, Phys. Reps. {\bf90} (1990) 307.

[31] P. Ginsparg and M. J. Perry, \npb 222 (1983) 245.

[32] G. W. Gibbons, \npb 207 (1982) 337;
\hb G. W. Gibbons and K. Maeda, \npb 298 (1988) 741;
\hb D. Garfinkle, G. T. Horowitz and A. Strominger,
\prd 43 (1991) 3140.

[33] I. Bars and D. Nemeschansky, \npb 348 (1991) 89;
\hb E. Witten, \prd 44 (1991) 314.

\ciao